\documentclass[prb,showpacs,twocolumn,preprintnumbers,
amsmath,groupedaddress,amssymb,superscriptaddress,floatfix]{revtex4}
\usepackage[dvips]{graphicx}
\usepackage[latin1]{inputenc}
\usepackage{bm}
\usepackage{color}
\begin{document}
\draft

\hyphenation{a-long}

\title{AC susceptibility investigation of vortex dynamics\\ in
nearly-optimally doped REFeAsO$_{1-x}$F$_{x}$ superconductors (RE =
La, Ce, Sm)}

\author{G.~Prando}
\email[E-mail: ]{giacomo.prando@unipv.it} \affiliation{Department of
Physics ``A. Volta,'' University of Pavia-CNISM, I-27100 Pavia,
Italy} \affiliation{Department of Physics ``E. Amaldi,'' University
of Roma Tre-CNISM, I-00146 Roma, Italy}
\author{P.~Carretta}
\affiliation{Department of Physics ``A. Volta,'' University of
Pavia-CNISM, I-27100 Pavia, Italy}
\author{R.~De Renzi}
\affiliation{Department of Physics, University of Parma-CNISM,
I-43124 Parma, Italy}
\author{S.~Sanna}
\affiliation{Department of Physics ``A. Volta,'' University of
Pavia-CNISM, I-27100 Pavia, Italy}
\author{H.-J.~Grafe}
\affiliation{Leibniz-Institut für Festkörper- und Werkstoffforschung
(IFW) Dresden, D-01171 Dresden, Germany}
\author{S.~Wurmehl}
\affiliation{Leibniz-Institut für Festkörper- und Werkstoffforschung
(IFW) Dresden, D-01171 Dresden, Germany}
\author{B.~Büchner}
\affiliation{Leibniz-Institut für Festkörper- und Werkstoffforschung
(IFW) Dresden, D-01171 Dresden, Germany}


\widetext

\begin{abstract}
Ac susceptibility and static magnetization measurements were
performed in the nearly-optimally doped LaFeAsO$_{0.9}$F$_{0.1}$ and
CeFeAsO$_{0.92}$F$_{0.08}$ superconductors, complementing earlier
results on SmFeAsO$_{0.8}$F$_{0.2}$ [Phys. Rev. {\bf B 83}, 174514
(2011)]. The magnetic field -- temperature phase diagram of the
mixed superconducting state is drawn for the three materials,
displaying a sizeable reduction of the liquid phase upon increasing
$T_{c}$ in the range of applied fields ($H \leq 5$ T). This result
indicates that SmFeAsO$_{0.8}$F$_{0.2}$ is the most interesting
compound among the investigated ones in view of possible
applications. The field-dependence of the intra-grain depinning
energy $U_{0}$ exhibits a common trend for all the samples with a
typical crossover field value ($2500$ Oe $\lesssim H_{cr} \lesssim
5000$ Oe) separating regions where single and collective depinning
processes are at work. Analysis of the data in terms of a simple
two-fluid picture for slightly anisotropic materials allows to
estimate the zero-temperature penetration depth $\lambda_{ab}(0)$
and the anisotropy parameter $\gamma$ for the three materials.
Finally, a sizeable suppression of the superfluid density is deduced
in a $s^{\pm}$ two-gap scenario.
\end{abstract}


\maketitle

\narrowtext

\section{Introduction}

Almost four years after the discovery of high-temperature
superconductivity in Fe-based pnictides,\cite{Kam08} several
questions are still open both on fundamental aspects and on possible
technological applications. No clear and exhaustive explanations for
the precise pairing mechanism have been given yet. Incontrovertible
features of isotopic effects cannot entirely rule out a partial role
of the lattice in the coupling process,\cite{Liu09} but it is
essential to stress that $T_{c}$'s values are indeed too high to
allow only a conventional pairing to be at work.\cite{Boe08} Strong
evidences for a multi-band $s^{\pm}$ scenario similar to that
proposed for magnesium di-boride\cite{Gon02} have been reported,
together with theoretical support,\cite{Bus10,Umm11} from a number
of techniques in materials belonging to the 1111
family,\cite{Hun08,Dag09,Lee09,Wey10} to 122 compounds\cite{Kha09}
and to 11 chalcogenides.\cite{Kha10}

At the same time, it is still not clear whether these novel
compounds will be helpful as valid technological tools. The answer
to these problems will essentially come from specific measurements
of critical current densities in relation to the features of the
grain boundaries. Detailed studies of the practical applicability of
granular superconductors, anyway, cannot leave aside precise
investigations of the fundamental intrinsic properties of the
materials. In this respect, the determination of the so-called
irreversibility line is of the utmost importance. Such line, in
fact, delimits the region in the magnetic field - temperature phase
diagram where the dissipationless property of the thermodynamical
superconducting phase is preserved even in the presence of a partial
penetration of magnetic field inside the material. In a previous
work on a powder sample of optimally-doped SmFeAsO$_{0.8}$F$_{0.2}$
(see Ref. \onlinecite{Pra11}) the determination of the intrinsic
irreversibility line was performed by means of the ac susceptibility
technique. Parameters like the superconducting critical temperature
$T_{c}$ are anyway well known to be strongly dependent on the
considered RE ion.\cite{Joh10} The question of whether (and how) the
properties of the phase diagram of the flux lines and of the
relative irreversibility lines too are indeed dependent on the RE
ion is then of extreme importance.

In this paper we report on the phase diagram of the flux lines in
three powder samples of REFeAsO$_{1-x}$F$_{x}$ superconductors (RE =
La, Ce, Sm) under conditions of nearly-optimal doping. In
particular, much attention is devoted to the determination of the
irreversibility line and its dependence on the RE ion. The features
of pinning mechanisms and, in particular, of the characteristic
depinning energy barriers are investigated in detail within a
thermally-activated framework. Results are interpreted and analyzed
by distinguishing two magnetic field regimes characterized by single
and collective depinning processes. Together with a simple two-fluid
model and a proper normalization, this allows us to make
experimental data collapse on the same curve, a feature indicative
of a common underlying mechanism independent on the precise
material. Reliable estimates of the zero-temperature penetration
depth $\lambda_{ab}(0)$ and of the dependence of the anisotropy
parameter $\gamma$ on the RE ion are finally given, together with
some interpretations of the observed behaviour in terms of a
two-band $s^{\pm}$ model. Data relative to the Sm-based sample have
already been presented in Ref. \onlinecite{Pra11} and will be
reported also here for the sake of clarity and completeness.

\section{Experimentals. Aspects of DC magnetization and AC
susceptibility} Powder samples of REFeAsO$_{1-x}$F$_{x}$ (with RE =
La, Ce, Sm and nominal F$^{-}$ contents $x = 0.1$, $0.08$ and $0.2$,
respectively) were synthesized as described in previous
works.\cite{Pra11,Pra11b,Kon09,Zhu08} Static magnetization $M_{dc}$
and ac susceptibility $\chi_{ac}$ measurements were performed by
means of a Quantum Design MPMS-XL7 SQUID magnetometer and of a
MPMS-XL5 SQUID susceptometer, respectively. In the latter case a
small alternating magnetic field $H_{ac}$ with frequency $\nu_{m}$
is superimposed to a much higher static magnetic field $H$.
Measurements were always performed in field-cooled (FC) conditions
with $H_{ac} = 0.0675 - 1.5$ Oe parallel to $\mu_{0}H$, which varied
up to 5 T, while $\nu_{m}$ ranged from $37$ to $1488$ Hz.

\begin{figure}[htbp]
\vspace{6.6cm} \includegraphics{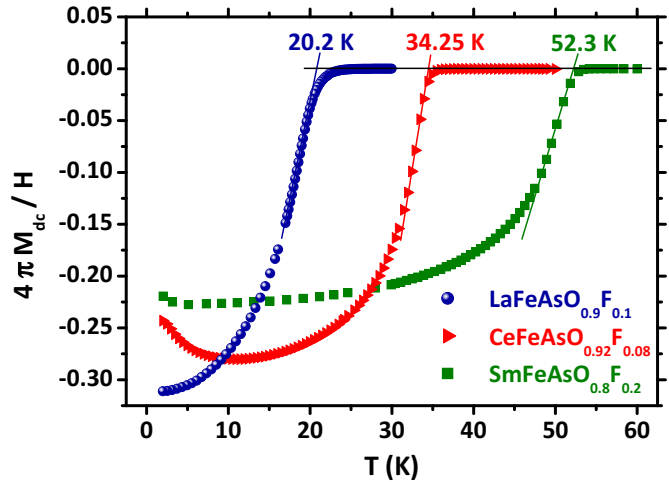} \caption{\label{FigMeissner}(Color online)
$M_{dc}/H$ vs. $T$ curves (volume units) in FC conditions at $H = 5$
Oe. The estimated $T_{c}(0)$ values are indicated in correspondence
of the diamagnetic onsets.}
\end{figure}

$M_{dc}/H$ vs. temperature ($T$) raw data obtained under FC
conditions at $H = 5$ Oe are shown in Fig. \ref{FigMeissner}. For
the sake of clarity the curves have been reported after subtracting
slight spurious contributions, leaving data above the
superconducting onset at a constant zero offset value. $T_{c}(0)$ is
defined as the critical temperature $T_{c}$ for $H \rightarrow 0$
Oe. Such values are obtained as the zero intercept of a linear
extrapolation of data below the diamagnetic onset\cite{Pra11} and
reported in Fig. \ref{FigMeissner} in correspondence of the
diamagnetic onsets of the three samples. The saturation absolute
value of the diamagnetic signals can be evaluated around $\left(0.27
\pm 0.05\right)$ in $1/4\pi$ units. These values are typical for
fully-superconducting powder samples in measurements under FC
conditions, the observed reduction originating from the sample's
geometrical properties and morphology. In the case of the
CeFeAsO$_{0.92}$F$_{0.08}$ sample, a sizeable paramagnetic
contribution from the Ce sublattice can be clearly discerned already
at such low value of magnetic field, mainly due to the high value of
the Ce$^{3+}$ magnetic moment. In fact, the fitting procedure to the
experimental data at several values of $H$ already described in a
previous work\cite{Pra11} yields to $\mu_{Ce} \simeq 2.1 \; \mu_{B}$
and $\mu_{Sm} \simeq 0.3 \; \mu_{B}$ (raw data not shown).

Curves reported in Fig. \ref{FigMeissner} show quite sharp
superconducting transitions. The slight roundness of the onset may
be due to several reasons, both intrinsic (for instance the
influence of superconducting fluctuations near $T_{c}$, as already
discussed in Ref. \onlinecite{Pra11b}) and extrinsic (slight
chemical inhomogeneity of the F$^{-}$ doping ions, distribution of
the geometrical size of grains). The signal, moreover, comes from
differently oriented grains in the powder sample and also this fact
may contribute to a broadening of the transition. The corresponding
powder-averaged upper critical field
\begin{equation}\label{EqPowderAverage}
    \langle H_{c2}(T)\rangle_{pwd} \simeq \frac{2}{3} H_{c2, H
    \parallel ab}(T) + \frac{1}{3} H_{c2, H \parallel c}(T)
\end{equation}
was deduced for the three samples by examining the field dependence
of $T_{c}$, as reported in Ref. \onlinecite{Pra11}.

\begin{figure*}[htbp]
\vspace{6.6cm} \includegraphics{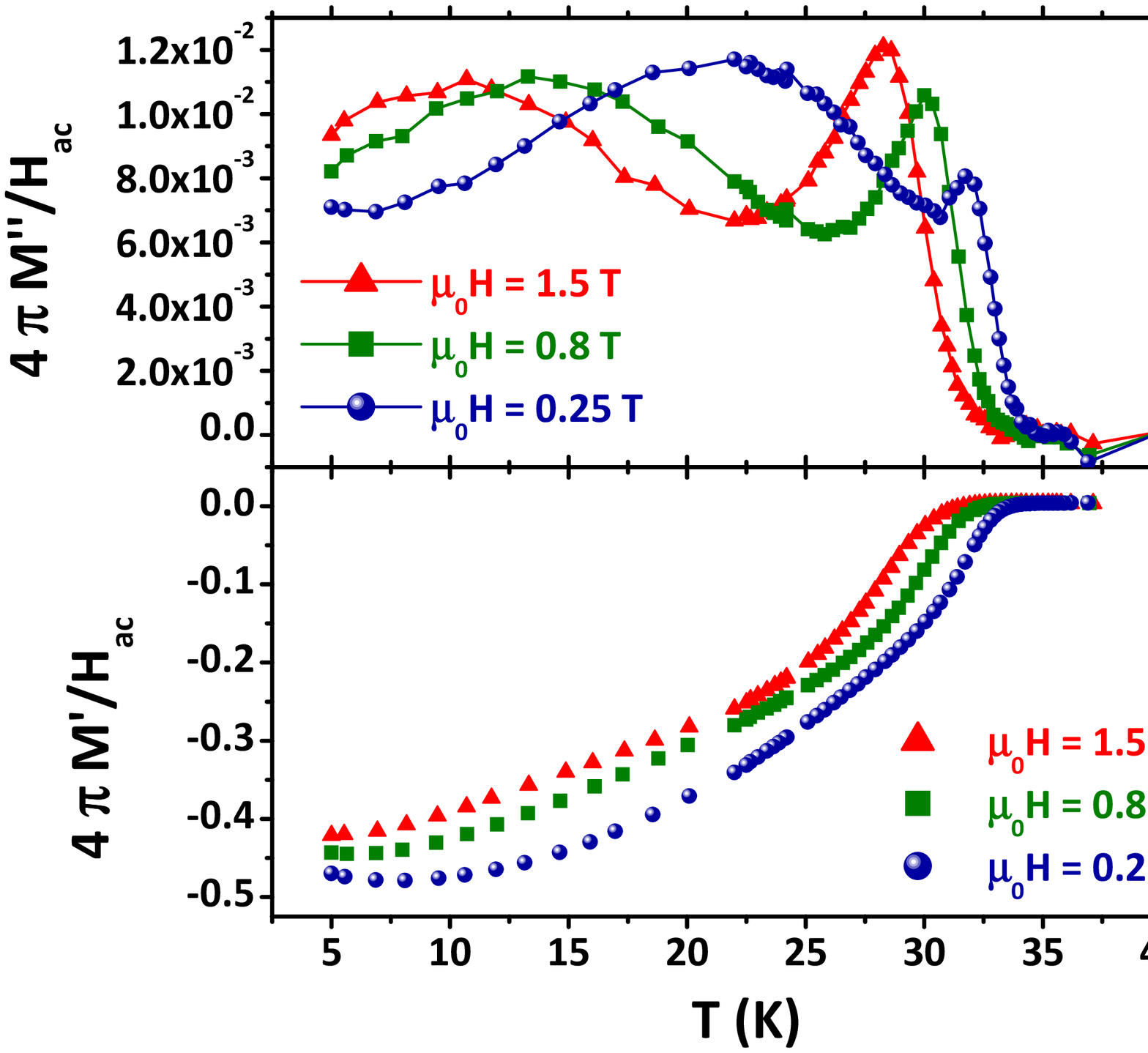} \hspace*{1.2cm}
\includegraphics{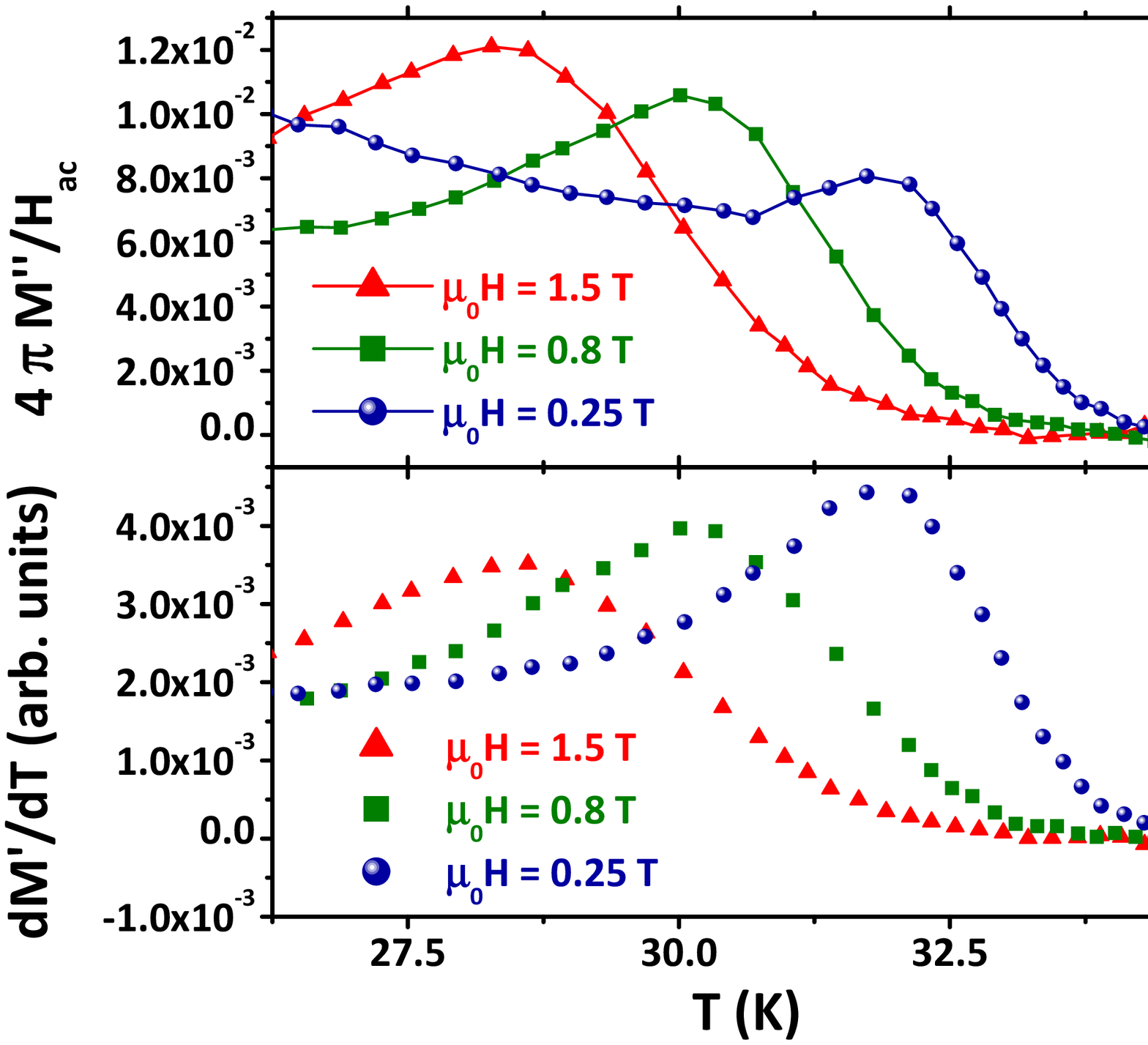} \caption{\label{FigDatiRawCe} (Color online) Left:
$\chi_{ac}$ vs. $T$ raw data for CeFeAsO$_{0.92}$F$_{0.08}$ ($H_{ac}
= 1.5$ Oe and $\nu_{m} = 478$ Hz) at different $H$ values. The
imaginary and the real components are shown in the upper and lower
panels, respectively (volume units). Right: zoom of the onset region
of the curves already reported in the left panels relative to
$\chi_{ac}^{\prime\prime}$ and to the first derivative of
$\chi_{ac}^{\prime}$ with respect to $T$ (upper and lower panels,
respectively. See text for further details).}
\end{figure*}

In Fig. \ref{FigDatiRawCe} some typical $\chi_{ac}$ vs. $T$ curves
for CeFeAsO$_{0.92}$F$_{0.08}$ are displayed (raw data of
SmFeAsO$_{0.8}$F$_{0.2}$ have been reported previously\cite{Pra11}).
Raw data for LaFeAsO$_{0.9}$F$_{0.1}$ are qualitatively similar to
those in Fig. \ref{FigDatiRawCe} and are not presented. The
$\chi_{ac}^{\prime}$ vs. $T$ curves can be described in terms of a
mixed-state shielding response with some degree of distortion
occurring in those $T$ regions where sizeable contributions to the
imaginary component appear. $\chi_{ac}^{\prime\prime}$ vs. $T$
curves are composed of two main peaked contributions: a narrow peak
appears just below the diamagnetic onset while a much broader one is
present at low $T$. This is quite a common phenomenology in
superconducting powder samples.\cite{Nik89,Gom97} The high-$T$ peak
is generally associated with the power absorption due to losses
inside the single grains, while the broad low-$T$ peak can be
associated with the generation of weak Josephson-like links among
the different grains.

The three main features shown in Fig. \ref{FigDatiRawCe}, namely the
diamagnetic onset in $\chi_{ac}^{\prime}$ and the two peaks in
$\chi_{ac}^{\prime\prime}$, are strongly shifted to lower $T$ on
increasing $H$ for all the samples. The $H$-dependence of the
diamagnetic onset in $\chi_{ac}^{\prime}$, in particular, is much
more marked than in the case of the diamagnetic onset in $M_{dc}/H$
vs. $T$ curves. The described behaviour can be directly associated
to vortex dynamics and to the precise features of the
irreversibility line.\cite{Pra11,Mal88,Tin88,VdB91,VdB93,Zhe94} When
dealing with an electromagnetic wave impinging on a type II
superconducting material, in fact, one has to carefully take into
consideration typical spatial penetration lengthscales. The general
treatment of the problem has been considered in several
papers.\cite{Bra91,Cof92,Pro03,Pro11} In the presence of vortices,
in particular, the overall penetration depth $\lambda_{ac}$ for the
radiation can be generally expressed as the quadrature sum of
$\lambda_{L}$ (representing the London penetration depth of the
superconductor) and $\lambda_{C}$ (the so-called Campbell
penetration depth), namely
\begin{equation}\label{EqFullPen}
    \lambda_{ac} = \sqrt{\lambda_{L}^{2} + \lambda_{C}^{2}}.
\end{equation}
$\lambda_{C}$ can be generally expressed as\cite{Pro11}
\begin{equation}\label{EqCamPen}
    \lambda_{C} = \sqrt{\frac{\Phi_{0}H}{4 \pi \alpha}}
\end{equation}
where the Labusch parameter $\alpha$ mimics the curvature of the
potential well associated with the pinning centers in a harmonic
approximation. In other terms, $\alpha$ quantifies the average
elastic restoring force density of the pinning centers acting on the
flux lines (FLs).

The case of high effectiveness of the pinning mechanisms ($\alpha
\rightarrow \infty$) ideally corresponds to a condition where FLs
are completely fixed. This condition is, by definition, the
so-called glassy phase of FLs where typically irreversible processes
develop.\cite{Yes96} By qualitatively considering Eqs.
\eqref{EqFullPen} and \eqref{EqCamPen} under these circumstancies,
one notices that the penetration of the electromagnetic wave is only
governed by the London penetration depth $\left(\lambda_{ac}
\rightarrow \lambda_{L}\right)$. As a result, the electromagnetic
wave is shielded by the superconductor leading to a diamagnetic
response in $\chi_{ac}^{\prime}$. On the other hand, in the opposite
case of completely ineffective pinning the Labusch constant can be
considered as a vanishing quantity. In this case, the FLs are in the
so-called liquid state and are substantially free to move generating
dissipation. The condition $\alpha \rightarrow 0$ yields to
$\lambda_{C} \rightarrow \infty$ or, at least, to $\lambda_{C} \gg
R$ in the case of a powder sample ($R$ is the typical grain size)
and no shielding can be detected even in the presence of a robust
thermodynamical superconducting phase. As a result,
$\chi_{ac}^{\prime} = 0$ even if $M_{dc}/H$ already takes negative
diamagnetic values. The onset of the diamagnetic response of the
material in $\chi_{ac}^{\prime}$ vs. $T$ curves can then be
interpreted as a crossover between the two described phases of the
FLs and its $H$-dependence is a good choice in order to define the
irreversibility line.

A slightly different criterion for determining the irreversibility
line can be formulated by considering the intrinsic dissipative
response inside the grains.\cite{Pra11,VdB91,Pal90,Emm91,Tin91} In
particular, by now considering the $\chi_{ac}^{\prime\prime}$ vs.
$T$ curves, one can denote by $T_{p}$ the position of the
intra-grain maximum which is typically found just few-K below the
diamagnetic onset in $\chi_{ac}^{\prime}$. This peak can be
interpreted as arising from a resonating absorption of energy when
the frequency of the radiation matches the inverse characteristic
relaxation time $1/\tau_{c}$ of the vortices in the pinning
potential dip, namely
\begin{equation}\label{EqResonatingAbsorptionAtTP}
    2 \pi \nu_{m} \tau_{c}|_{{}_{T = T_{p}}} = 1.
\end{equation}
The $H$-dependence of $T_{p}$ can then be chosen in order to
describe the irreversibility line and in the following we will be
referring to this as the $\chi_{ac}^{\prime\prime}$-criterion.
Within a Debye-like relaxation framework, $T_{p}$ almost coincides
with the characteristic temperature of the corresponding peak in the
derivative of $\chi_{ac}^{\prime}$ with respect to $T$. This was
experimentally verified in all the three investigated
samples,\cite{Pra11} as explicitly shown only in the case of
CeFeAsO$_{0.92}$F$_{0.08}$ in the right panel of Fig.
\ref{FigDatiRawCe}.

\section{Phase diagrams and depinning
energy barriers: main results} 
\begin{figure}[htbp]
\vspace{6.6cm} \includegraphics{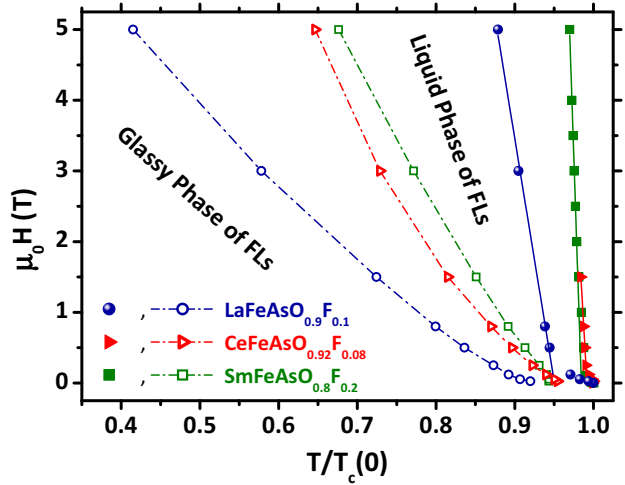} \caption{\label{FigPhDiag}(Color online)
Phase diagrams for the FLs in the investigated samples. Data for
SmFeAsO$_{0.8}$F$_{0.2}$ have already been reported in Ref.
\onlinecite{Pra11}. Open symbols track the irreversibility line as
deduced from data taken at $\nu_{m} = 37$ Hz (dashed-dotted lines
are guides for the eye). Full symbols track $\langle
H_{c2}(T)\rangle_{pwd}$ (continuous lines are linear fits).}
\end{figure}

The FLs phase diagrams for the three
examined samples are shown in Fig. \ref{FigPhDiag} where both
$\langle H_{c2}(T)\rangle_{pwd}$ (full symbols) and the
irreversibility lines (opened symbols) are plotted as a function of
the reduced temperature $t \equiv T/T_{c}(0)$.

\begin{center}
\begin{table}[b!]
    \caption{Critical temperature $T_{c}(0)$ for $H \rightarrow 0$
    Oe and correlation length $\langle\xi(0)\rangle_{pwd}$ for
    $T \rightarrow 0$ K as deduced from Eqs. \eqref{EqWHH} and
    \eqref{EqHc2GL}.}\label{TabHc2Parameters} 
    \begin{tabular}{c||c|c}%
        \toprule%
        Sample & \;\;\; $T_{c}(0)$ (K) \;\;\;
        & \; $\langle\xi(0)\rangle_{pwd}$ (Å) \;\\
        \toprule%
        SmFeAsO$_{0.8}$F$_{0.2}$ & $52.3 \pm 0.25$ &
        $11.0 \pm 0.1$\\
        CeFeAsO$_{0.92}$F$_{0.08}$ & $34.25 \pm 0.25$ &
        $15.7 \pm 0.3$\\
        LaFeAsO$_{0.9}$F$_{0.1}$ & $20.2 \pm 0.25$ &
        $26.3 \pm 0.5$\\
        \toprule%
    \end{tabular}
\end{table}
\end{center}

$\langle H_{c2}(T)\rangle_{pwd}$ vs. $t$ curves as determined from
$M_{dc}$ data clearly display linear trends as a function of $t$. At
low-$t$ values, anyway, a slight upper curvature can be detected for
all the samples, a feature possibly associated with two-band
superconductivity.\cite{Gur03} Data relative to
CeFeAsO$_{0.92}$F$_{0.08}$ are limited to the low-$H$ values due to
the dominant paramagnetic contribution arising from the Ce$^{3+}$
sublattice which fully covers the superconducting response for
$\mu_{0}H \gtrsim 2$ T. The slope of a linear fit to the data allows
to estimate the upper critical field $\langle
H_{c2}(0)\rangle_{pwd}$ extrapolated at $T = 0$ K under the
simplified assumption of single-band superconductivity through the
Werthamer, Helfand and Hohenberg (WHH) relation\cite{Wer66}
\begin{equation}\label{EqWHH}
    \langle H_{c2}(0)\rangle_{pwd} \simeq 0.7 \times T_{c}(0)
    \left|\frac{d \langle H_{c2}\rangle_{pwd}}
    {d T}\right|_{T \simeq T_{c}(0)}.
\end{equation}
An overall correlation is observed between the gradual increase of
$T_{c}(0)$ and a corresponding increase in the slope value and,
accordingly, in the extrapolated $\langle H_{c2}(0)\rangle_{pwd}$.
This, in turn, leads to a steady decrease of the extrapolated value
of the powder-averaged Ginzburg-Landau (GL) coherence length
$\langle \xi(0)\rangle_{pwd}$ at $T = 0$ K, calculated from the
relation\cite{Tin96,DeG99} (see Tab. \ref{TabHc2Parameters})
\begin{equation}\label{EqHc2GL}
    \langle H_{c2}(0)\rangle_{pwd} \equiv \frac{\Phi_{0}}
    {2\pi\langle\xi(0)\rangle_{pwd}^{2}}.
\end{equation}

Results from $\chi_{ac}$ data will now be considered. The
$\chi_{ac}^{\prime\prime}$-criterion was chosen for the
determination of the irreversibility line. Data relative to the
derivative of $\chi_{ac}^{\prime}$ were always analyzed due to their
more favourable signal-to-noise ratio. A dependence of the $T_{p}$
value on $\nu_{m}$ was detected at all the values of the applied
$H$. For this reason, data for the lowest accessible value $\nu_{m}
= 37$ Hz were chosen in order to draw the irreversibility lines in
Fig. \ref{FigPhDiag}. It is possible to observe that the extension
of the liquid phase of the flux lines is progressively reduced by
the increase of $T_{c}$ in the explored $H$ range. This observation
makes Sm-based materials rather interesting in view of possible
technological applications.

\begin{figure}[b!]
\vspace{6.6cm} \includegraphics{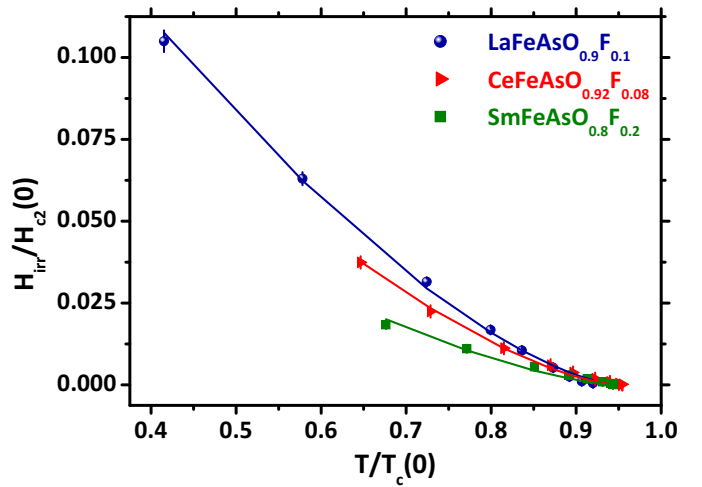}
\caption{\label{FigNormalizedPhaseDiagFLs}(Color online)
Representation of the irreversibility lines relative to the
investigated samples (already reported in the phase diagrams, see
Fig. \ref{FigPhDiag}) where $H_{irr}(T)$ has been normalized with
respect to the relative value of $\langle H_{c2}(0)\rangle_{pwd}$.
Continuous lines are best-fits according to Eq. \eqref{EqPowLaw}.}
\end{figure}
In order to better compare the behaviour of the three samples from a
fundamental point of view, the irreversibility lines reported in
Fig. \ref{FigPhDiag} are presented again in Fig.
\ref{FigNormalizedPhaseDiagFLs} after normalizing the field values
by $\langle H_{c2}(0)\rangle_{pwd}$. In all the three samples the
irreversibility line can be described by means of a power-law
function
\begin{equation}\label{EqPowLaw}
    \frac{H}{\langle H_{c2}(0)\rangle_{pwd}} \propto
    \left[0.95 - \frac{T}{T_{c}(0)}\right]^{3/2}
\end{equation}
characterized by the exponent $\beta = 3/2$ (see the continuous
lines in Fig. \ref{FigNormalizedPhaseDiagFLs}). This is a typical
result in high-$T_{c}$ superconductors,\cite{Pra11,Mal88,Yes88} even
if slightly different functional form have been reported, for
instance, in the case of Ba(Fe$_{1-x}$Co$_{x}$)$_{2}$As$_{2}$
single-crystals.\cite{Pro08} Here the value $0.95$ (common for all
the samples) phenomenologically accounts for the discrepancies at
low magnetic fields when defining the irreversibility line from the
$\chi_{ac}^{\prime\prime}$-criterion. It should be noticed that,
when plotted on this different scale for the different samples, the
extension of the liquid phase as a function of the RE ion shows a
trend opposite to what displayed in Fig. \ref{FigPhDiag}. This
interesting feature will be recalled and discussed later in Sect.
\ref{SectDiscussion}.

\begin{figure}[htbp]
\vspace{6.6cm} \includegraphics{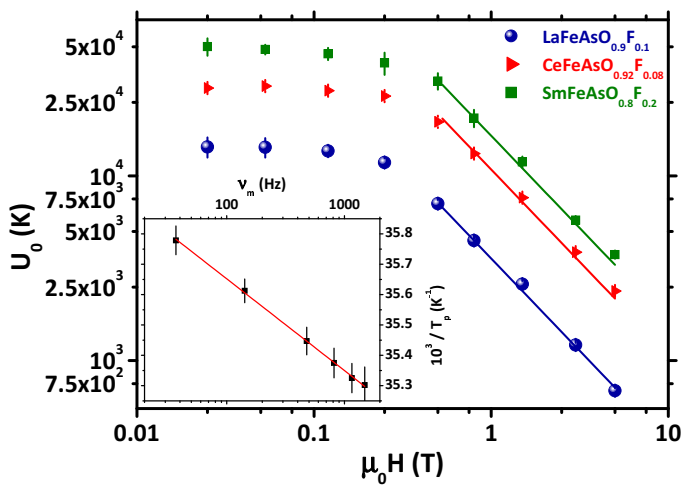} \caption{\label{FigUDepinning}(Color
online) Main panel: $H$-dependence of the depinning energy barriers
$\langle U_{0}(H)\rangle_{pwd}$ in the three investigated samples.
Data relative to SmFeAsO$_{0.8}$F$_{0.2}$ have already been reported
in a previous work.\cite{Pra11} Continuous lines are best-fits to
data according to a simple $1/H$ dependence. Inset: logarithmic
$\nu_{m}$-dependence of $1/T_{p}$ in CeFeAsO$_{0.92}$F$_{0.08}$
($H_{ac} = 1.5$ Oe while $\mu_{0}H = 1.5$ T). The continuous line is
a best-fit function according to Eq. \eqref{EqLogDep}.}
\end{figure}

By now focussing on the $\nu_{m}$-dependence of $T_{p}$ one can
notice that, similarly to what observed in
SmFeAsO$_{0.8}$F$_{0.2}$,\cite{Pra11} the quantity $1/T_{p}$
displays a logarithmic dependence on $\nu_{m}$. This is clearly
shown in the inset of Fig. \ref{FigUDepinning} for
CeFeAsO$_{0.92}$F$_{0.08}$ ($H_{ac} = 1.5$ Oe, $\mu_{0}H = 1.5$ T)
even if the phenomenology is well verified for all the samples at
all the $H$ values. In particular, data can always be fitted within
a thermally-activated framework by the expression
\begin{equation}\label{EqLogDep}
    \frac{1}{T_{p}\left(\nu_{m}\right)}
    = -\frac{1}{\langle U_{0}(H) \rangle_{pwd}}
    \ln\left(\frac{\nu_{m}}{\nu_{0}}\right)
\end{equation}
(see the fitting function in the inset of Fig. \ref{FigUDepinning}).
One can recognize that the logarithmic behaviour of $1/T_{p}$ is
mainly controlled by the powder-averaged fitting parameter $\langle
U_{0}(H)\rangle_{pwd}$, playing the role of an effective depinning
energy barrier in a thermally-activated flux creep model. The
parameter $\nu_{0}$ in Eq. \eqref{EqLogDep} represents an
intra-valley characteristic frequency associated with the motion of
the vortices around their equilibrium position in the pinning
centers.

The results of the fitting procedure to $T_{p}\left(\nu_{m}\right)$
data according to Eq. \eqref{EqLogDep} have been reported in the
main panel of Fig. \ref{FigUDepinning} for all the samples. A
strongly-marked $H$-dependence of $\langle U_{0}(H)\rangle_{pwd}$
can clearly be discerned. Remarkably, a common trend is displayed
for all the samples, making it possible to guess a common underlying
mechanism. Beyond an overall sizeable reduction of $\langle
U_{0}(H)\rangle_{pwd}$ with increasing $H$, a sharp crossover (at
field values $2500$ Oe $\lesssim H_{cr} \lesssim 5000$ Oe common to
all the samples) between two qualitatively different $\langle
U_{0}(H)\rangle_{pwd}$ vs. $H$ regimes is observed for each sample.
At low fields $H < H_{cr}$ the depinning energy $\langle
U_{0}(H)\rangle_{pwd}$ is found to be only slightly dependent on
$H$, while for values $H > H_{cr}$ a trend $\langle
U_{0}(H)\rangle_{pwd} \propto 1/H$ can be discerned. The $1/H$
regime for depinning energies $\langle U_{0}(H)\rangle_{pwd}$ was
indeed observed in different superconducting materials by means of
several techniques ranging from magnetoresistivity\cite{Pal90} to ac
magnetometry itself\cite{Emm91} and nuclear magnetic
resonance.\cite{Rig98} This feature has already been reported in the
SmFeAsO$_{0.8}$F$_{0.2}$ sample\cite{Pra11} and, as it will be
discussed later on, it can be justified in terms of pinning effects
on a single FL propagating among bundles of entangled FLs. In this
framework, the crossover between the two different trends of
$\langle U_{0}(H)\rangle_{pwd}$ vs. $H$ shown in Fig.
\ref{FigUDepinning} can be interpreted as the transition from a
basically single-flux line response at low $H$ values to a
collective response of vortices for $H > 5000$ Oe. The saturated
low-$H$ values for the depinning energy barriers are typically
$\langle U_{0}(H)\rangle_{pwd} \sim 10^{4}$ K. Such high values are
in agreement with what reported from magnetoresistivity measurements
in SmFeAsO$_{0.85}$ (see Ref. \onlinecite{Lee10}) and in
Ba$_{1-x}$K$_{x}$Fe$_{2}$As$_{2}$ single-crystals, even if in the
latter case a much weaker $H$-dependence was observed.\cite{Wan10}
These results are clearly indicative of strong intrinsic pinning of
the vortex lines. Features of strong-pinning mechanisms have also
been deduced in single-crystals of PrFeAsO$_{1-x}$ and
NdFeAsO$_{1-x}$F$_{x}$, confirming that the observed behaviour is
intrinsic in REFeAsO$_{1-x}$F$_{x}$ materials.\cite{VdB10}

It must be noticed that, since the variation of $1/T_{p}$ vs.
$\nu_{m}$ is very modest (see the vertical scale in the inset of
Fig. \ref{FigUDepinning}), $\langle U_{0}(H)\rangle_{pwd}$ is not
only determined at a fixed $H$ but almost in isothermal conditions
as well. By referring to the inset of Fig. \ref{FigUDepinning}, in
fact, one can recognize that $T_{p}(\nu_{m})$ is varying on a range
of some tenths of K degree, as already observed in the case of the
Sm-based sample.\cite{Pra11} This is a great advantage of the ac
susceptibility technique if compared, for instance, to
magnetoresistivity measurements where $\langle
U_{0}(H)\rangle_{pwd}$ is estimated from an activated-like fit to
data over a range of several tens of K degrees.\cite{Pal90,Lee10}
One can then notice that the average temperature $T^{*}$
characterizing the variation of $T_{p}$ over the considered
$\nu_{m}$ range is intrinsically positioned over the irreversibility
line, so that the energy barrier should more correctly be referred
to as $\langle U_{0}(H^{*},T^{*})\rangle_{pwd}$, where
$(H^{*},T^{*})$ are the points on the $H$ -- $T$ phase diagram
belonging to the irreversibility line.

\section{Analysis of the results}

The observed behaviour can be explained by means of the
phenomenological GL theory\cite{Tin96} and by referring to the model
of single-vortex pinning by atomic impurities (like, for instance,
ionic substitutions like O$^{2-}$/F$^{-}$ or O$^{2-}$
vacancies).\cite{Mal88,Tin88,Yes88} Due to the very small values of
the coherence lengths $\sim 10$ Å, in fact, local defects on the
atomic scale can be considered as strongly efficient pinning centers
for FLs.\cite{Lee10} In this framework the energy $U_{0}$ is related
to the FLs features only and does not depend on the precise pinning
mechanisms.\cite{Tin88} Moreover, at strong $H$-values the high
density of FLs gives rise to entangled bundles of vortices around
the central one physically bound to the atomic
defect.\cite{Mal88,Tin88,Yes88} Accordingly, the characteristic
energy $U_{0}$ can be directly linked to the geometrical properties
of the flux line lattice and, in particular, to the typical volume
$V$ of the correlated vortex lines. The following phenomenological
expression can be envisaged (where $U_{0}$ is expressed in
K)\cite{Mal88,Tin88,Yes88}
\begin{equation}\label{EqVolumeGen}
    U_{0}(T) = \left\{\frac{H_{c}^{2}(T)}{8\pi}\right\} \frac{V}{k_{B}}
\end{equation}
where the term between curly brackets quantifies the $T$-dependence
of the superconducting condensation energy density. Two limiting
cases can be considered.

On the one hand, by gradually increasing $H$ the density of vortices
steadily increases and, accordingly, correlations among vortices
increase too. A crossover to a regime where the depinning process
leads to a collective response of an increasing number of vortices
is then expected. In the simplified scenario of a square Abrikosov
lattice of vortices, the quantity $d = \sqrt{\Phi_{0}/H}$ estimates
the mutual distance among nearest neighbouring FLs. One can then
assume that the correlations roughly extend over a cylindric volume
whose radius is given by $d$. Namely one has $V \simeq
\pi\xi(T)\Phi_{0}/H$, the characteristic size along the third
dimension being determined by the coherence length.\cite{Yes88}
After the substitution in Eq. \eqref{EqVolumeGen}, considerations on
the powder-averaging making it possible to obtain a more convenient
form in order to describe the experimental data. In particular, the
following relation holds under a two-fluid approximation (see
Appendix \ref{AppendixHF} for details)
\begin{eqnarray}\label{EqU0NormDefHF}
    \langle U_{0}(H^{*},T^{*})\rangle_{pwd} &=&
    \frac{\Phi_{0}^{5/2}}{96\sqrt{2}\pi^{3/2}k_{B}}
    \frac{g(t^{*})\sqrt{\langle
    H_{c2}(0)\rangle_{pwd}}}{H^{*}}\times\nonumber\\ &\times&
    \frac{1}{f_{1}(\gamma)\lambda_{ab}^{2}(0)}
\end{eqnarray}
where $g(t)$ is a function of $t$ and $f_{1}(\gamma)$ is a function
of the anisotropy parameter $\gamma = \xi_{ab}/\xi_{c} =
\lambda_{c}/\lambda_{ab}$.

On the other hand, at low enough magnetic fields the typical volume
$V$ will be no longer sensitive to the presence of several flux
lines but it will only be a function of the typical lengths of the
single pinned vortex. As a consequence, one can consider a different
cylindric volume with radius $\delta \xi(T)$. Here the heuristic
parameter $\delta$ is introduced in order to grant a continuous
crossover between the high- and the low-$H$ regimes.\cite{Yes88}
Thus, again choosing the coherence length as the characteristic size
also along the third dimension, the following relation follows
\begin{eqnarray}\label{EqU0NormDefLF}
    \langle U_{0}(H^{*},T^{*})\rangle_{pwd} & = &
    \frac{\Phi_{0}^{5/2}\delta^{2}}{192\sqrt{2}\pi^{5/2}k_{B}} \frac{h(t^{*})}
    {\sqrt{\langle H_{c2}(0)\rangle_{pwd}}}
    \times\nonumber\\ & \times &
    \frac{1}{f_{2}(\gamma)\lambda_{ab}^{2}(0)}.
\end{eqnarray}
Here $h(t)$ is a function of $t$ and $f_{2}(\gamma)$ is a function
of the anisotropy parameter (details relative to the derivation of
Eq. \eqref{EqU0NormDefLF} can be found in Appendix
\ref{AppendixLF}).

The two resulting expressions show that in both the $H$-regimes
$\lambda_{ab}(0)$ can be simply expressed as a function of the
experimentally-accessible quantity $\langle
U_{0}(H^{*},T^{*})\rangle_{pwd}$. One should consider that the
continuity of $\lambda_{ab}(0)$ must clearly be assured at the
crossover field.\cite{Yes88} The free parameters for the three
different samples (namely, the anisotropy parameter $\gamma_{RE}$
and the parameter $\delta_{RE}$), anyway, make the fulfilment of
this requirement quite arbitrary. Some other criterion from the
analysis of experimental data should be formulated in order to
derive some relations among the parameters, reducing as much as
possible the degree of arbitrariness. This can be done by the
examination of the FLs phase diagram and by taking into
consideration the role of thermal fluctuations. As a final result of
the procedure, as reported in detail in Appendix
\ref{AppendixLinking}, one finds that
\begin{equation}\label{EqConstraints}
    \gamma_{Sm}^{3/2}\delta_{Sm}^{2} \simeq 1.6
    \gamma_{Ce}^{3/2}\delta_{Ce}^{2} \simeq 3
    \gamma_{La}^{3/2}\delta_{La}^{2}.
\end{equation}
The linking up the different sets of data can now be performed under
the fulfilment of the constraints reported in Eq.
\eqref{EqConstraints}. As a starting point of the procedure, a
reasonable value for one of the $\gamma_{RE}$ parameters should be
taken as fixed. By referring to typical data reported in
literature,\cite{Pal09} in particular, it will be assumed that
$\gamma_{Sm} = 5$. Next, data relative to the two different regimes
(high- and low-$H$) in SmFeAsO$_{0.8}$F$_{0.2}$ are linked up by
setting a proper value of $\delta_{Sm}$. The procedure is repeated
also for Ce- and La-based samples, where the starting guesses for
the parameters $\gamma_{RE}$ and $\delta_{RE}$ must be modified till
the fulfilment of Eq. \eqref{EqConstraints}.

\section{Discussion}\label{SectDiscussion}

\begin{figure}[htbp]
\vspace{6.6cm} \includegraphics{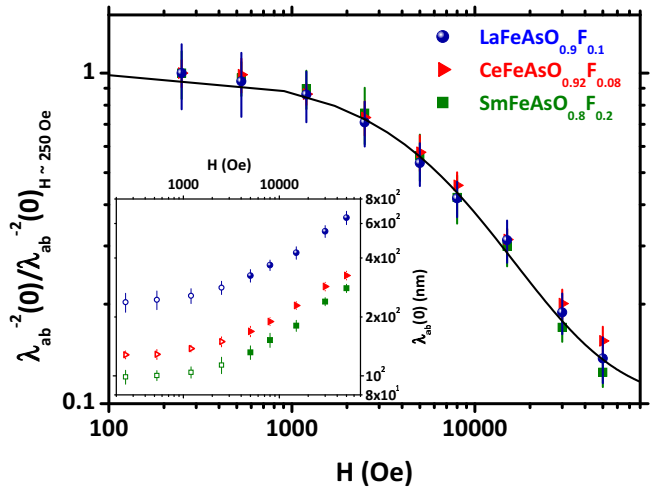}
\caption{\label{FigNormU0HFCommon}(Color online) Main panel:
graphical representation of the collapse of experimental data
represented in Fig. \ref{FigUDepinning} after a proper normalization
according to the value of $1/\lambda_{ab}^{2}(0)$ estimated at $250$
Oe. The continuous line is a best-fit to data according to Eq.
\eqref{EqSuppressionSuperfluidDensity}. Inset: $H$-dependence of the
$\lambda_{ab}(0)$ values for the three investigated samples. Open
(full) symbols are relative to data from the low-$H$ (high-$H$)
regime.}
\end{figure}
As reported in the inset of Fig. \ref{FigNormU0HFCommon}, the
linking procedure of data in the two opposite limits described by
Eqs. \eqref{EqU0NormDefHF} and \eqref{EqU0NormDefLF} under the
constraints reported in Eq. \eqref{EqConstraints} allows one to
deduce a sizeable $H$-dependence of $\lambda_{ab}(0)$ itself. The
saturation values in the limit $H \rightarrow 0$ Oe have been
reported in the last column of Tab. \ref{TabConstraints} compared to
the results of $\mu^{+}$SR measurements on the same samples (raw
data not shown), together with the results for $\gamma_{RE}$ and
$\delta_{RE}$. The presented results of $\mu^{+}$SR measurements are
in good agreement with results on different samples characterized by
slightly different stoichiometries.\cite{Kha08,Lue08} At the same
time, the estimates of $\lambda_{ab}(0)$ for
CeFeAsO$_{0.92}$F$_{0.08}$ and SmFeAsO$_{0.8}$F$_{0.2}$ by means of
$\chi_{ac}$ are consistent with the values from $\mu^{+}$SR
measurements within a systematic factor $\sim 2$. This discrepancy
could possibly be accounted for by assuming that the correlations
among vortices in the high-$H$ regime are actually extended over a
bigger volume. In particular, it is immediate to realize that the
choice of $2d$ for the cylindric volume, also implying a rescaling
$\delta \rightarrow 2\delta$, leads to a complete agreement of
$\chi_{ac}$ data with $\mu^{+}$SR ones.

\begin{table}[b!]
\caption{Parameters $\delta$, $\gamma$ and $\lambda_{ab}(0)$
(estrapolated for $H \rightarrow 0$ Oe) for the three samples
resulting from the constrained linking procedure of data relative to
the two distinct $H$-regimes (see text for details). Values of
$\lambda_{ab}(0)$ as obtained from $\mu^{+}$SR measurements have
been reported in the last column (raw data not shown in the text).}
\label{TabConstraints} 
\begin{center}
\begin{tabular}{c||c|c|c|c}%
\toprule%
Sample & $\;\;\; \gamma \;\;\;$ & $\;\;\; \delta \;\;\;$ &
$\lambda_{ab}(0)$ (nm) & $\lambda_{ab}(0)$ (nm)\\
{} & {} & {} & from $\chi_{ac}$ & from $\mu^{+}$SR\\
\toprule%
LaFeAsO$_{0.9}$F$_{0.1}$ & 3.2 & 9.5 & 235 $\pm$ 15 & $\diagup$\\
CeFeAsO$_{0.92}$F$_{0.08}$ & 4.25 & 10.5 & 128 $\pm$ 6 & 260 $\pm$ 15 \\
SmFeAsO$_{0.8}$F$_{0.2}$ & 5 & 11.75 & 98 $\pm$ 8 & 200 $\pm$ 15 \\
\toprule%
\end{tabular}
\end{center}
\end{table}

The observed behaviour of $\delta$ as a function of the RE ion can
be qualitatively understood in terms of a compensation of the
opposite trend in $\langle\xi\rangle_{pwd}$. The trend in the
modification of the value of $\gamma$, on the other hand, can be
correlated with the increase in the size of the liquid region in the
phase diagram reported in Fig. \ref{FigNormalizedPhaseDiagFLs}. One
can deduce that the increase in $\gamma$ possibly leads to the
enhancement of 2-dimensional fluctuations, much more effective than
the higher-dimensional ones in extending the liquid region of the
phase diagram. It should be remarked that a similar trend in La- and
Sm-based samples was reported in literature, even if the measured
absolute values were considerably higher.\cite{Jar08}

More interestingly, a scaling of $\lambda_{ab}^{-2}(0)$ is displayed
in the main panel of Fig. \ref{FigNormU0HFCommon} showing a clearly
common $H$-dependence after a proper normalization with respect to
the value $\lambda_{ab}^{-2}(0)$ taken at $H = 250$ Oe. This feature
possibly demonstrates the existence of a common material-independent
underlying behaviour. It should also be remarked that the quantity
$\lambda_{ab}^{-2}(0)$ is directly associated with the superfluid
density $n_{s}$ of the superconducting state.\cite{Tin96,DeG99}
This, in turn, implies that $n_{s}$ is partially suppressed by $H$
values much lower than $H_{c2}(0)$, a characteristic fingerprint of
multi-gap superconductivity.\cite{Wey10,Kha08b} By assuming a
$H$-dependence typical of $s$-like bands\cite{Son00} one can derive
the following phenomenological fitting function
\begin{equation}\label{EqSuppressionSuperfluidDensity}
    n_{s}(H) = n_{s1} + n_{s2}(H) = n_{s1} + \frac{1-n_{s1}}
    {1+\left(\frac{H}{H_{0}}\right)^{2}}.
\end{equation}
The fitting results show that the main contribution to the
superfluid density $\left(1-n_{s1}\right) \simeq 0.9$ comes from the
weakest band that is suppressed by a typical magnetic field $H_{0}
\simeq 1.25 \times 10^{4}$ Oe. This quite unphysical result should
possibly be associated with the high degree of approximation
associated with the two-fluid model and the relative $T$-dependence
of the characteristic lengths (see Eqs.
\eqref{EqTwoFluidTDependence} later in Appendix \ref{AppendixHF}).
Our results are anyway in good qualitative agreement with previous
reports obtained by means of torque magnetometry and local muon spin
spectroscopy on an optimally-doped Sm-based
superconductor.\cite{Wey10} In that case from $\mu^{+}$SR it was
possible to deduce a weaker $H$-dependence, eventually saturating at
$\mu_{0}H \sim 1.5$ T and involving the suppression of just $25$\%
of the overall superfluid density.

\section{Conclusions}

The phase diagram of the flux lines in three powder samples of
REFeAsO$_{1-x}$F$_{x}$ superconductors (RE = La, Ce, Sm) under
conditions of nearly-optimal doping was investigated by means of ac
susceptibility measurements. The irreversibility line has been
estimated for the three samples, showing that in the accessible
range of magnetic field the extension of the reversible liquid
region is lowered with increasing $T_{c}$. This aspect could be of
extreme interest in view of possible technological applications of
these materials. The $H$-dependence of the characteristic depinning
energy barriers was investigated in a thermally-activated framework.
Results were interpreted by distinguishing two regimes of magnetic
field characterized by single and collective depinning processes,
allowing us to make experimental data collapse on the same
temperature trend indicative of a common underlying mechanism
independent on the precise material. Reliable estimates of the
zero-temperature penetration depth $\lambda_{ab}(0)$ and of the
RE-dependence of the anisotropy parameter $\gamma$ are finally
given. Some interpretations of the observed behaviour in terms of a
two-band $s^{\pm}$ model are proposed even if a refinement of the
employed two-fluid model is in order.

\section*{Acknowledgements}

R. Khasanov, G. Lamura and A. Rigamonti are gratefully acknowledged
for stimulating discussions. H. Stummer, C. Malbrich, S.
Müller-Litvanyi, R. Müller, K. Leger, J. Werner and S. Pichl are
acknowledged for assistance in the sample preparation. The work at
the IFW Dresden was supported by the Deutsche Forschungsgemeinschaft
through Grant No. BE1749/12 the Priority Program SPP1458 (Grant No
BE1749/13). S. W. acknowledges support by DFG under the Emmy-Noether
program (Grant No. WU595/3-1).

\appendix

\section{Derivation of the relations between
$\langle U_{0}(H^{*},T^{*})\rangle_{pwd}$ and $\lambda_{ab}(0)$ in
the two different $H$-regimes}

\subsection{Strongly correlated vortices. Depinning of bundles of flux
lines}\label{AppendixHF} Let's consider Eq. \eqref{EqVolumeGen} in
the limit of high-$H$, here rewritten for convenience
\begin{equation}
    U_{0}(H,T) = \frac{H_{c}^{2}(T)}{8 k_{B}}
    \frac{\xi(T)\Phi_{0}}{H},
\end{equation}
as the starting point. By means of the GL relation for the flux
quantum $\Phi_{0} = 2\sqrt{2}\pi H_{c}(T)\lambda(T)\xi(T)$, one can
write down an explicit expression for $U_{0}$ in different
conditions of orientation of the magnetic field as follows
\begin{eqnarray}
    U_{0, H\parallel ab}(H,T) & = & \frac{\Phi_{0}^{3}}{64\pi^{2}k_{B}H}
    \frac{1}{\sqrt{\xi_{ab}(T)\xi_{c}(T)}\lambda_{ab}(T)\lambda_{c}(T)}
    \nonumber\\
    U_{0, H\parallel c}(H,T) & = & \frac{\Phi_{0}^{3}}{64\pi^{2}k_{B}H}
    \frac{1}{\xi_{ab}(T)\lambda_{ab}^{2}(T)}.
\end{eqnarray}
The expressions $\xi_{ab}/\xi_{c}=\lambda_{c}/\lambda_{ab} = \gamma$
hold for the different typical lengths in anisotropic
superconductors, where $\gamma$ has already been defined in the text
as the anisotropy parameter. By referring to Eq.
\eqref{EqPowderAverage} it is possible to perform a powder-like
average of $U_{0}$ as
\begin{eqnarray}
    \langle U_{0}(H,T)\rangle_{pwd} & \equiv & \frac{2}{3} U_{0, H\parallel
    ab}(T) + \frac{1}{3} U_{0, H\parallel c}(T) = {}\\
    & =  &
    \left(\frac{\sqrt{\gamma}+2}{3\gamma^{3/2}}\right)
    \frac{\Phi_{0}^{3}}{64\pi^{2}k_{B}H}
    \frac{1}{\xi_{c}(T)\lambda_{ab}^{2}(T)}\nonumber.
\end{eqnarray}
In a simple two-fluid model, the $T$-dependence of $\xi_{c}(T)$ and
$\lambda_{ab}(T)$ can be taken as
\begin{equation}\label{EqTwoFluidTDependence}
    \xi_{c}(T) = \xi_{c}(0) \frac{\sqrt{1-t^{4}}}
    {1-t^{2}},\;\;\;
    \lambda_{ab}(T) = \lambda_{ab}(0) \frac{1}
    {\sqrt{1-t^{4}}}
\end{equation}
where $t$ has already been defined as the reduced temperature $t
\equiv T/T_{c}(0)$. It should be considered that, as already
recalled, the estimate of $\langle U_{0}(H,T)\rangle_{pwd}$ is
performed by definition at the values $(H^{*},T^{*})$ delimiting the
irreversibility line. After the definition of the function $g(t)
\equiv \left(1-t^{2}\right)\sqrt{1-t^{4}}$ it is then possible to
write
\begin{eqnarray}\label{EqU0NormHF}
    \frac{\langle U_{0}(H^{*},T^{*})\rangle_{pwd}}{g(t^{*})} &=&
    \left(\frac{\sqrt{\gamma}+2}{3\gamma^{3/2}}\right)
    \frac{\Phi_{0}^{3}}{64\pi^{2}k_{B}H^{*}}\times\nonumber\\
    &\times& \frac{1}{\xi_{c}(0)\lambda_{ab}^{2}(0)}.
\end{eqnarray}
The quantity $\xi_{c}(0)$ can be independently derived by measuring
the magnetic field dependence of the superconducting transition
temperature $T_{c}$ by means of the relations reported in Eqs.
\eqref{EqWHH} and \eqref{EqHc2GL} after proper considerations about
the powder-average procedures. The two limiting configurations
$H\parallel ab$ and $H\parallel c$ lead to the formulas
\begin{eqnarray}
    H_{c2, H\parallel ab}(T) & = & \frac{\Phi_{0}}{2\pi
    \xi_{ab}(T)\xi_{c}(T)},\nonumber\\
    H_{c2, H\parallel c}(T) & = & \frac{\Phi_{0}}{2\pi
    \xi_{ab}^{2}(T)}.
\end{eqnarray}
Following Eq. \eqref{EqPowderAverage} it is possible to deduce that
the experimentally-accessible quantity $\langle\xi(0)\rangle_{pwd}$,
already defined in Eq. \eqref{EqHc2GL}, is linked to $\xi_{c}(0)$ by
the relation
\begin{equation}\label{EqRelationCsiCCsiPowderAveraged}
    \langle\xi(0)\rangle_{pwd} = \xi_{c}(0) \left(\frac{\sqrt{3}\gamma}
    {\sqrt{2\gamma+1}}\right).
\end{equation}
Coming back to Eq. \eqref{EqU0NormHF}, one can substitute Eqs.
\eqref{EqRelationCsiCCsiPowderAveraged} and \eqref{EqHc2GL} to
obtain
\begin{eqnarray}\label{EqU0NormDefHFbis}
    \frac{\langle U_{0}(H^{*},T^{*})\rangle_{pwd}}{g(t^{*})\sqrt{\langle
    H_{c2}(0)\rangle_{pwd}}} &=&
    \frac{\Phi_{0}^{5/2}}{96\sqrt{2}\pi^{3/2}k_{B}}\times\nonumber\\ &\times&
    \frac{1}{f_{1}(\gamma)\lambda_{ab}^{2}(0)}\frac{1}{H^{*}}
\end{eqnarray}
having defined the function of the anisotropy parameter
\begin{equation}
    f_{1}(\gamma) = \frac{1}{\sqrt{3}}
    \frac{\sqrt{2\gamma^{2}+\gamma}}{\sqrt{\gamma} + 2}.
\end{equation}
Eq. \eqref{EqU0NormDefHFbis} is equivalent to Eq.
\eqref{EqU0NormDefHF}.

\subsection{Weakly correlated vortices. Depinning of single flux
lines}\label{AppendixLF} Let's now consider Eq. \eqref{EqVolumeGen}
in the limit of low-$H$, here rewritten for convenience
\begin{equation}
    U_{0}(T) = \frac{H_{c}^{2}(T)}{8 k_{B}}
    \delta^{2}\xi^{3}(T).
\end{equation}
Again by means of the GL relation for the flux quantum $\Phi_{0} =
2\sqrt{2}\pi H_{c}(T)\lambda(T)\xi(T)$ it is possible to explicit
the expressions for $U_{0}$ in the different cases of the
orientation of the system with respect to the magnetic field as
\begin{eqnarray}
    U_{0, H\parallel ab}(T) & = & \frac{\Phi_{0}^{2}\delta^{2}}{64\pi^{2}k_{B}}
    \frac{\sqrt{\xi_{ab}(T)\xi_{c}(T)}}{\lambda_{ab}(T)\lambda_{c}(T)}\nonumber\\
    U_{0, H\parallel c}(T) & = & \frac{\Phi_{0}^{2}\delta^{2}}{64\pi^{2}k_{B}}
    \frac{\xi_{ab}(T)}{\lambda_{ab}^{2}(T)}.
\end{eqnarray}
Similarly to what performed in the previous Appendix concerning
strongly correlated vortices, the $\gamma$ factor can be introduced
and $\lambda_{ab}(T)$, $\xi_{c}(T)$ can be left as independent
quantities. By considering that the estimate of $\langle
U_{0}(H,T)\rangle_{pwd}$ is performed along the irreversibility line
$(H^{*},T^{*})$, by employing Eq. \eqref{EqTwoFluidTDependence} and
after a powder-like average of the energy barrier one can write
\begin{eqnarray}\label{EqU0NormLF}
    \frac{\langle U_{0}(H^{*},T^{*})\rangle_{pwd}}{h(t^{*})} &=&
    \left(\frac{\gamma^{2}+2\sqrt{\gamma}}{3\gamma}\right)
    \frac{\Phi_{0}^{2}\delta^{2}}{64\pi^{2}k_{B}}\times\nonumber\\
    &\times& \frac{\xi_{c}(0)}{\lambda_{ab}^{2}(0)}
\end{eqnarray}
where $h(t) \equiv \left(1+t^{2}\right)\sqrt{1-t^{4}}$. By again
expressing $\xi_{c}(0)$ in terms of $\langle H_{c2}(0)\rangle_{pwd}$
through Eq. \eqref{EqRelationCsiCCsiPowderAveraged} and by means of
Eq. \eqref{EqHc2GL}, it is finally possible to deduce the following
expression
\begin{eqnarray}\label{EqU0NormDefLFbis}
    \frac{\langle U_{0}(H^{*},T^{*})\rangle_{pwd}\sqrt{\langle
    H_{c2}(0)\rangle_{pwd}}}{h(t^{*})} & = &
    \frac{\Phi_{0}^{5/2}\delta^{2}}{192\sqrt{2}\pi^{5/2}k_{B}}
    \times\nonumber\\ & \times &
    \frac{1}{f_{2}(\gamma)\lambda_{ab}^{2}(0)}.
\end{eqnarray}
In the previous expression, $f_{2}(\gamma)$ is defined as
\begin{equation}
    f_{2}(\gamma) =
    \frac{\sqrt{3}\gamma^{2}}{\left(\gamma^{2}+2\sqrt{\gamma}\right)
    \sqrt{2\gamma + 1}}.
\end{equation}
Eq. \eqref{EqU0NormDefLFbis} is equivalent to Eq.
\eqref{EqU0NormDefLF}.

\section{Linking procedure of data in
the two different $H$-regimes} \label{AppendixLinking}

In this Appendix the problem of the continuity at the crossover
field $H_{cr}$ of the $\lambda_{ab}(0)$ vs. $H$ data, obtained by
means of Eqs. \eqref{EqU0NormDefHF} and \eqref{EqU0NormDefLF}, will
be considered. In particular, as already stated in the text, some
constraints on the variability of the six parameters $\gamma_{RE}$
and $\delta_{RE}$ (RE = La, Ce, Sm) should be fixed in order to
reduce as much as possible the degree of arbitrariness of the
procedure of data-linking.

At this aim, it is convenient to introduce the Ginzburg-Levanyuk
number $\widetilde{Gi}_{3D}(0)$ as\cite{Lar05}
\begin{equation}\label{EqGinzburgLandauNumberZero}
    \widetilde{Gi}_{3D}(0) = \frac{1}{2}
    \left[\frac{8\pi^{2}k_{B}T_{c}(0)\lambda^{2}(0)}
    {\Phi_{0}^{2}\xi(0)}\right]^{2}
\end{equation}
quantifying the extension of the region of the $H-T$ phase diagram
where thermal fluctuations are sizeable and significantly affect the
physics of the system. The $H$-dependence of
$\widetilde{Gi}_{3D}(0)$ is given by\cite{Lar05}
\begin{equation}\label{EqHDepGiNumber}
    Gi_{3D}(0,H) = \left[\frac{2H
    \sqrt{\widetilde{Gi}_{3D}(0)}}{H_{c2}(0)}\right]^{2/3}.
\end{equation}
In fact, one can assume that the position of the irreversibility
line is mainly governed by the amount of thermal fluctuations in the
system. As a consequence, $Gi_{3D}(0,H)$ is expected to be directly
involved in the analytic expression relative to the irreversibility
line itself. One hint at the correctness of this picture is possibly
given by the similarity between the characteristic exponents
observed in Eqs. \eqref{EqPowLaw} and \eqref{EqHDepGiNumber}.
Similar considerations, moreover, have already been proposed in
literature concerning the thermodynamical melting line (see, in
particular, Sections IV and V of Ref. \onlinecite{Bla94} and
references therein. In that case, anyway, the considered exponent is
$\beta = 2$). The following expression for the irreversibility line
can then be considered
\begin{equation}
    1 - \frac{T^{*}}{T_{c}(0)} = \frac{Gi_{3D}(0,H^{*})}{K^{2/3}}
\end{equation}
where $K$ is an arbitrary proportionality factor. Together with Eq.
\eqref{EqHDepGiNumber}, this straightforwardly leads to
\begin{equation}\label{EqPhenExprPhDiag}
    \frac{2H^{*}\sqrt{\widetilde{Gi}_{3D}(0)}}{H_{c2}(0)} = K
    \left[1 - \frac{T^{*}}{T_{c}(0)}\right]^{3/2}
\end{equation}
Since all the sample-dependent quantities are already kept into
consideration by $\widetilde{Gi}_{3D}(0)$, it is reasonable to
assume $K$ as a sample-independent parameter. $K$ can typically be
interpreted in terms of microscopic properties of the
vortices.\cite{Bla94} In the present phenomenological model, anyway,
such microscopic interpretations are left aside.

In order to give a suitable description of the experimental data,
Eq. \eqref{EqPhenExprPhDiag} should be powder-averaged by
considering the criterion already presented in Eq.
\eqref{EqPowderAverage}. By considering the definition of
$\widetilde{Gi}_{3D}(0)$ reported in Eq.
\eqref{EqGinzburgLandauNumberZero}, one can write
\begin{eqnarray}
    \frac{\sqrt{\widetilde{Gi}_{3D}(0)}}{H_{c2}(0)}
    & = & \frac{1}{\sqrt{2}}
    \frac{8\pi^{2}k_{B}T_{c}(0)\lambda_{ab}(0)\lambda_{c}(0)}
    {\Phi_{0}^{2}\sqrt{\xi_{ab}(0)\xi_{c}(0)}}
    \frac{2\pi \xi_{ab}(0)\xi_{c}(0)}{\Phi_{0}}\nonumber\\
    \frac{\sqrt{\widetilde{Gi}_{3D}(0)}}{H_{c2}(0)}
    & = & \frac{1}{\sqrt{2}}
    \frac{8\pi^{2}k_{B}T_{c}(0)\lambda_{ab}^{2}(0)}
    {\Phi_{0}^{2}\xi_{ab}(0)}
    \frac{2\pi \xi_{ab}^{2}(0)}{\Phi_{0}}
\end{eqnarray}
for the two field orientations $H\parallel ab$ and $H\parallel c$,
respectively. By again performing the powder-average procedure and
after introducing the anisotropy parameter through the relations
$\xi_{ab}/\xi_{c}=\lambda_{c}/\lambda_{ab} = \gamma$, by referring
to Eqs. \eqref{EqHc2GL} and \eqref{EqRelationCsiCCsiPowderAveraged}
one obtains the following expression
\begin{eqnarray}\label{EqPowderAverageIrreversibilityLineFirst}
    \left\langle\frac{\sqrt{\widetilde{Gi}_{3D}(0)}}{H_{c2}(0)}
    \right\rangle_{pwd} &=& \frac{1}{f_{3}(\gamma)}
    \frac{8\pi^{2}k_{B}T_{c}(0)\lambda_{ab}^{2}(0)}
    {\sqrt{2}\Phi_{0}^{2}\xi_{c}(0)}\times\nonumber\\
    &\times& \frac{1}{\langle H_{c2}(0)\rangle_{pwd}}
\end{eqnarray}
where the function $f_{3}(\gamma)$ is defined as
\begin{equation}
    f_{3}(\gamma) = \frac{9\gamma^{2}}{\left(2\gamma^{3/2}+\gamma\right)
    \left(2\gamma+1\right)}.
\end{equation}

One should now consider that the definition of
$\widetilde{Gi}_{3D}(0)$ reported in Eq.
\eqref{EqGinzburgLandauNumberZero} does not account for any
$H$-dependence that is fully accounted for by Eq.
\eqref{EqHDepGiNumber}. Eq.
\eqref{EqPowderAverageIrreversibilityLineFirst} is then referred to
a $H = 0$ Oe condition and, as a consequence, the quantity
$\lambda_{ab}^{2}(0)/\xi_{c}(0)$ can be obtained by means of Eq.
\eqref{EqU0NormLF} (holding in the low-$H$ regime). By inserting the
resulting expression into Eq. \eqref{EqPhenExprPhDiag} one obtains
\begin{equation}\label{EqPowderAveragedIrreversibility}
    \frac{H^{*}}{\langle H_{c2}(0)\rangle_{pwd}} \simeq
    \widetilde{K}_{RE} \times \left[0.95 -
    \frac{T^{*}}{T_{c}(0)}\right]^{3/2}
\end{equation}
where
\begin{equation}\label{EqPowderAveragedIrreversibilityBis}
    \widetilde{K}_{RE} \equiv
    \left\{\frac{\langle U_{0}(H^{*},T^{*})\rangle_{pwd}}
    {T_{c}(0)h(t^{*})}\right\}_{RE}\times
    \frac{27\sqrt{2}} {\gamma_{RE}^{3/2}\delta_{RE}^{2}}K.
\end{equation}
As already discussed in the text, the factor $0.95$ in Eq.
\eqref{EqPowderAveragedIrreversibility} phenomenologically accounts
for the definition of the irreversibility line from the
$\chi_{ac}^{\prime\prime}$-criterion.

\begin{figure}[t!]
\vspace{6.6cm} \includegraphics{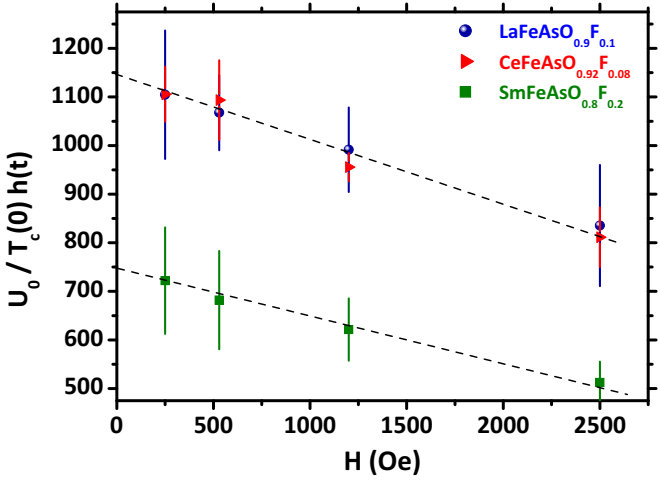}
\caption{\label{FigConstraints}(Color online) Graphical
representation of the quantity between square brackets in Eq.
\eqref{EqPowderAveragedIrreversibilityBis}. The dashed lines are
guides for the eye and they can be used to extrapolate the intercept
values back to $H = 0$ Oe.}
\end{figure}
The quantity between curly brackets in Eq.
\eqref{EqPowderAveragedIrreversibilityBis} can be experimentally
estimated from a linear extrapolation of the intercept values back
to $H = 0$ Oe (see Fig. \ref{FigConstraints}). One can then compare
the sample-dependent quantity $\widetilde{K}_{RE}$ with the
corresponding experimental quantities derived from the fitting
procedure shown in Fig. \ref{FigNormalizedPhaseDiagFLs}. Together
with the already cited assumption that $K$ is a sample-independent
quantity, this implies that some constraints can be put on the
variability of the parameters $\gamma_{RE}$ and $\delta_{RE}$. In
particular, one finds that
\begin{equation}
    \gamma_{Sm}^{3/2}\delta_{Sm}^{2} \simeq 1.6
    \gamma_{Ce}^{3/2}\delta_{Ce}^{2} \simeq 3
    \gamma_{La}^{3/2}\delta_{La}^{2},
\end{equation}
as reported in Eq. \eqref{EqConstraints}.



\end{document}